\documentclass{PoS}

\usepackage{wrapfig}

\usepackage{amsmath}
\usepackage{amsfonts,amsbsy}
\usepackage{amssymb}
\usepackage{appendix}
\usepackage{array}
\usepackage{multirow}
\def\be{\begin{equation}}
\def\ee{\end{equation}}
\def\bea{\begin{eqnarray}}
\def\eea{\end{eqnarray}}

\def\eq#1{{Eq.~(\ref{#1})}}
\def\fig#1{{Fig.~\ref{#1}}}

\def\be{\begin{equation}}
\def\ee{\end{equation}}
\def\bea{\begin{eqnarray}}
\def\eea{\end{eqnarray}}

\def\eq#1{{Eq.~(\ref{#1})}}
\def\fig#1{{Fig.~\ref{#1}}}

\def\beq{\begin{equation}}
\def\eeq{\end{equation}}
\def\bea{\begin{eqnarray}}
\def\eea{\end{eqnarray}}

\renewcommand{\vec}[1]{\boldsymbol{#1}}
\newcommand{\dif}{\mathrm{d}}

\title{Diffractive photoproduction of vector mesons at the LHC}

\ShortTitle{Diffractive photoproduction of vector mesons at the LHC}

\author{N\'estor Armesto\\
  Departamento de F\'isica de Part\'iculas and IGFAE, Universidade de Santiago de Compostela, 15782 Santiago de Compostela, Galicia, Spain \\
   E-mail: \email{Nestor.Armesto@usc.es}}
\author{\speaker{Amir H. Rezaeian}
\\
Departamento de F\'\i sica, Universidad T\'ecnica
Federico Santa Mar\'\i a, Avda. Espa\~na 1680,
Casilla 110-V, Valparaiso, Chile\\
and\\  
 Centro Cient\'\i fico Tecnol\'ogico de Valpara\'\i so (CCTVal), Universidad T\'ecnica
Federico Santa Mar\'\i a, Casilla 110-V, Valpara\'\i so, Chile\\
E-mail: \email{Amir.Rezaeian@usm.cl}}

\abstract{ We confront saturation-based results for diffractive $\psi(2s)$ and $\rho$  production at HERA
and $J/\psi$ photoproduction with all available data including recent ones from HERA, ALICE and LHCb, finding a good agreement.  We show that the $t$-distribution of differential cross-section of photoproduction of vector mesons offers a unique opportunity to  discriminate among saturation and non-saturation models. This is due to emergence of a pronounced dip  (or multiple dips) in the $t$-distribution of diffractive photoproduction of vector mesons at relatively large, but potentially accessible $|t|$ that can be traced back to the unitarity features of colour dipole amplitude in the saturation regime.  We provide  various predictions for exclusive (photo)-production of different vector mesons including the ratio of $\psi(2s)/J/\psi$ at HERA, the LHC and at future colliders. 
}

\FullConference{XXII. International Workshop on Deep-Inelastic Scattering and Related Subjects,\\
		28 April - 2 May 2014\\
		Warsaw, Poland}

\begin{document}

\section{Introduction}
Exclusive diffractive vector meson production provides a rich testing ground of many QCD novel properties  and probes the high-energy limit of QCD. There is strong theoretical evidence that QCD at high-energy (or small Bjorken-$x$) leads to a non-linear regime where gluon recombination or unitarity effects become important \cite{sg,mv}, resulting in saturation of parton densities in hadrons and nuclei.  An effective field theory describing the high-energy limit of QCD is the Color Glass Condensate (CGC) \cite{mv}. One of the advantage of the CGC framework is that it is possible to simultaneously describe other high-energy hadronic interactions in a regime not currently accessible to approaches that rely on collinear factorisation, see e.g. Refs.\,\cite{Albacete:2014fwa,dihadron,diphoton,fac} and references therein.

The LHCb and ALICE collaborations have recently released new data on $J/\psi$ photoproduction with photon-proton center-of-mass energies upto about 1.3 TeV \cite{lhcb,lhcbn}, the highest energy ever measured so far in this kind of reaction.  Alongside this, the recently released high-precision combined HERA data that were not available at the time of previous studies of  diffractive processes, provide extra important constraints on saturation models \cite{bcgc-new,ipsat-new}. The main purpose of this study is to confront the saturation based predictions with those recent data  from the LHC in order to examine the importance of the saturation effect.   We systematically study elastic diffractive production of different vector mesons $J/\psi$, $\psi(2s)$ and $\rho$  off protons and  investigate which vector meson production is more sensitive to saturation physics and what measurement can be potentially a better probe of the signal.  In particular, we study $\psi(2s)$ diffractive production by constructing the $\psi(2s)$ forward wave function via a fit to the leptonic decay, and provide various predictions for diffractive $\psi(2s)$ production as well as the ratio of $\psi(2s)/J/\psi$ at HERA and the LHC.  Below, we summarize a few key results,  the details can be found in Ref.\,\cite{na-VM}.

\vspace{-0.2cm}
\section{Exclusive diffractive processes in the colour-dipole formalism}
\vspace{-0.2cm}
 Similar to the case of the inclusive DIS process,  the scattering amplitude for the exclusive diffractive process $\gamma^*+p\to V+p$, with a final state vector meson $V=J/\psi, \psi(2s), \phi,\rho$  (or a real photon $V=\gamma$), can be written in terms of a convolution of the dipole amplitude $\mathcal{N}$ and the overlap of the wave functions of the photon and the exclusive final state particle (see \cite{bcgc-new,ipsat-new,na-VM} and references therein), 
\begin{equation} \label{am-i}
  \mathcal{A}^{\gamma^* p\rightarrow Vp}_{T,L} = \mathrm{2i}\,\int\!\dif^2\vec{r}\int\!\dif^2\vec{b}\int_0^1\!\dif{z}\;(\Psi_{V}^{*}\Psi)_{T,L}(r,z,m_f,M_V;Q^2)\;\mathrm{e}^{-\mathrm{i}[\vec{b}-(1-z)\vec{r}]\cdot\vec{\Delta}}\mathcal{N}\left(x,r,b\right),
\end{equation}
where $\vec{\Delta}^2=-t$ with $t$ being the squared momentum transfer. In this equation, $\mathcal{N}$ is the imaginary part of the forward $q\bar{q}$ dipole-proton scattering amplitude with transverse dipole size $r$ and impact parameter $b$. The parameter $z$ is the fraction of the light cone momentum of the virtual photon carried by the quark and $m_f$ denotes the mass of the quark with flavour $f$.  In \eq{am-i} summations over the quark helicities and over the quark flavour $f=u,d,s, c$ are implicit and the $\Psi_{V}^{*}\Psi$ is the forward overlap wave function of photon and vector meson. The differential cross-section of the exclusive diffractive processes can then be written in terms of the scattering amplitude as \cite{bcgc-new,ipsat-new,na-VM},
\begin{equation}
  \frac{\dif\sigma^{\gamma^* p\rightarrow Vp}_{T,L}}{\dif t} = \frac{1}{16\pi}\left\lvert\mathcal{A}^{\gamma^* p\rightarrow Vp}_{T,L}\right\rvert^2\;(1+\beta^2) R_g^{2},
  \label{vm}
\end{equation}
with
\bea \label{eq:beta} 
  \beta &=& \tan\left(\frac{\pi\delta}{2}\right), \hspace{0.5cm} R_g(\delta) = \frac{2^{2\delta+3}}{\sqrt{\pi}}\frac{\Gamma(\delta+5/2)}{\Gamma(\delta+4)}, \hspace{0.5cm}
\delta \equiv \frac{\partial\ln\left(\mathcal{A}_{T,L}^{\gamma^* p\rightarrow Vp}\right)}{\partial\ln(1/x)}. \
\eea 

For the forward $q\bar{q}$ dipole-proton scattering amplitude  in \eq{am-i}, we employ the impact-parameter dependent Color Glass Condensate (b-CGC) \cite{bcgc-new} and Saturation (IP-Sat) \cite{ipsat-new} dipole models which were recently updated with high precision HERA combined data. Both models incorporate key features of small-x physics properties and match smoothly to the perturbative QCD regime at large $Q^2$ for a given $x$.  Both models have been intensively applied to many reactions including heavy-ion collisions, see e.g. \cite{exp1,exp2}. In order to single out the implication of saturation effect, we compare our results with 1-Pomeron model \cite{na-VM} which corresponds to the leading-order pQCD expansion for the dipole amplitude in the colour-transparency region, as opposed to the saturation case. For the forward vector meson wave functions, we employ the boosted Gaussian wave-function with parameters determined from normalisation, the orthogonality conditions and a fit to the experimental leptonic decay width \cite{na-VM}.  Note that both $\Psi_{V}^{*}\Psi$ and  $\mathcal{N}$ are external input here which were constrained in other reactions than those considered here. The details of the overlap wavefunctions $\Psi_{V}^{*}\Psi$ and the dipole amplitudes $\mathcal{N}$ can be found in Ref\footnote{In the revised version of \cite{na-VM} (v3) we include the details of vector meson wavefunctions and show the values of their free parameters determined via a fit to leptonic decays.}.\,\cite{na-VM}.


\begin{figure}[t]       
\includegraphics[width=0.49\textwidth,clip]{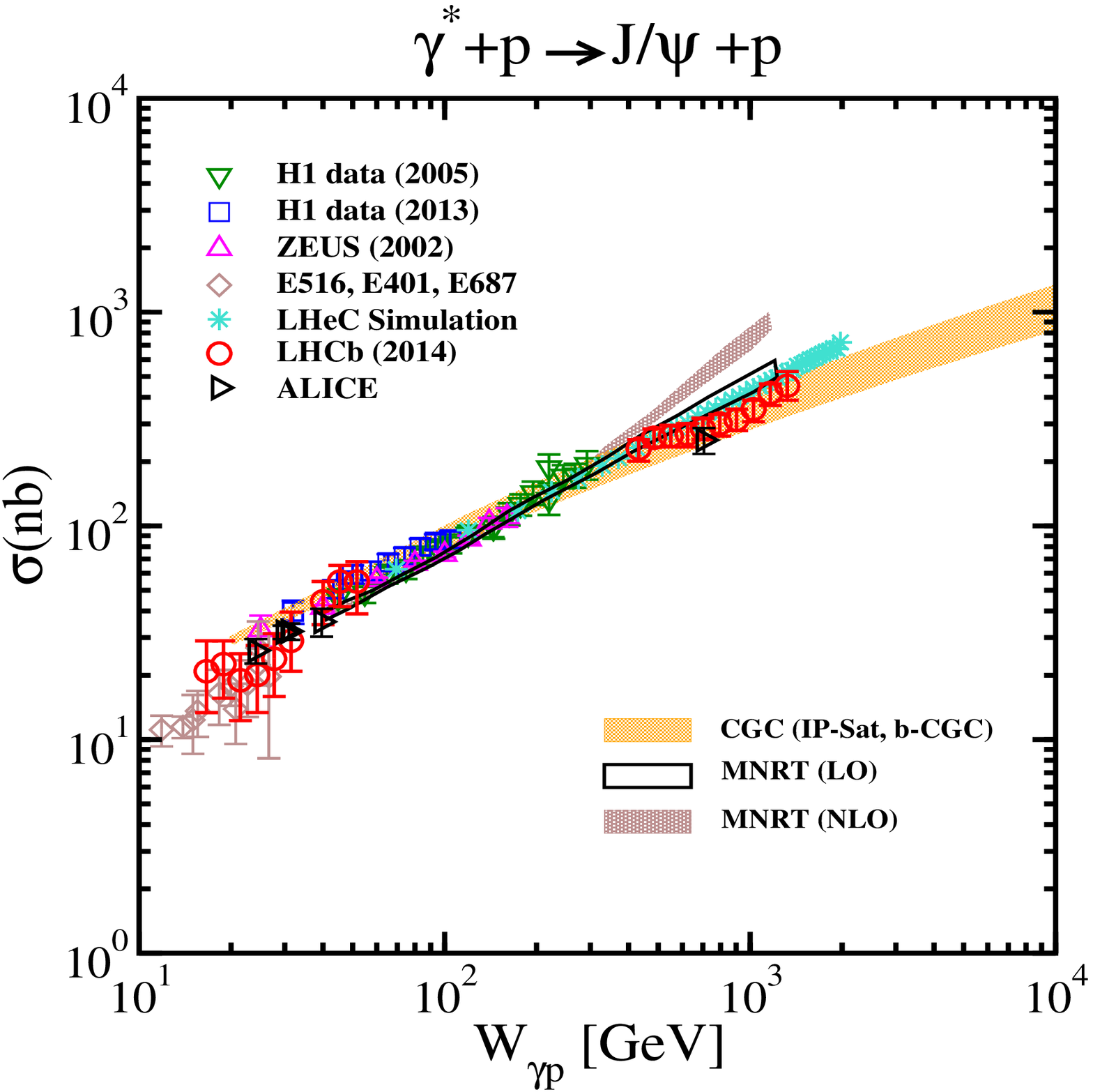}
\includegraphics[width=0.49\textwidth,clip]{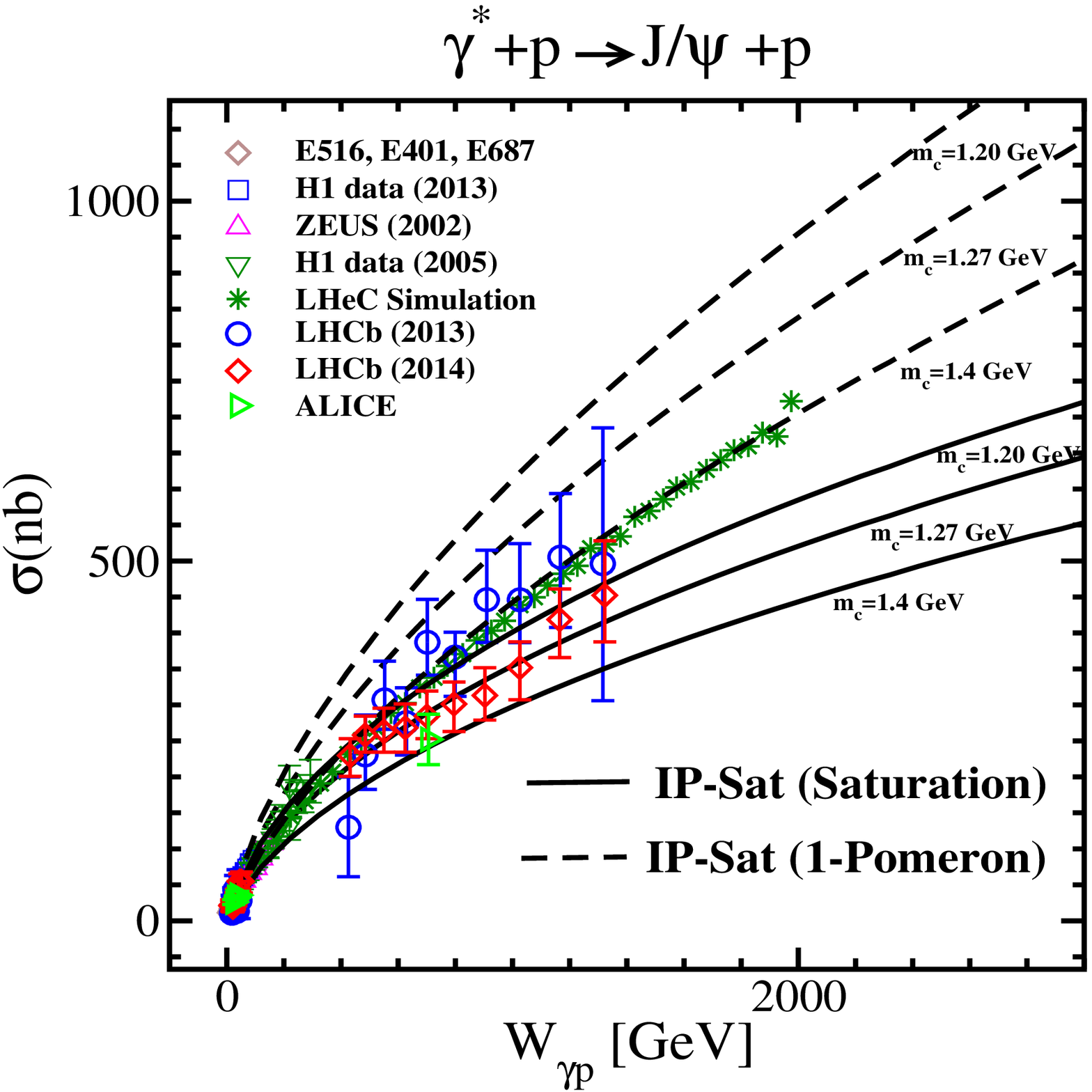}
\caption{Left: Total $J/\psi$ cross-section as a function of $W_{\gamma p}$, compared to results from the CGC/saturation (orange band) calculated from the b-CGC and the IP-Sat models \cite{bcgc-new,ipsat-new}. Right: Total $J/\psi$ cross-section as a function of $W_{\gamma p}$, compared to the results from the IP-Sat (saturation) and 1-Pomeron models with different charm mass $m_c$.The plots are taken from Ref.\,\cite{na-VM}.}
  \label{f-vw1}
\end{figure}   

In \fig{f-vw1}, we compare  the total $J/\psi$ cross-section as a function of centre-of-mass energy of the photon-proton system $W_{\gamma p}$, obtained from saturation models and from a pQCD approach at leading-order (LO) and next-to-leading-order (NLO) \cite{pqcd-j} with all available data from fixed target experiments to the recent ones from HERA, LHCb and ALICE.  We also show the LHeC pseudo-data obtained from a simulation based on a power-law extrapolation of HERA data.  The band labeled  "CGC" includes the saturation results obtained from the IP-Sat and the b-CGC models with the parameters of models constrained by the recent combined HERA data. Note that the LHCb data points in \fig{f-vw1} were not used for fixing the model parameters, and therefore our CGC results in \fig{f-vw1} at high energy can be considered as predictions. Also note that diffractive $J/\psi$ production is sensitive to the charm quark mass at low $Q^2$. This is because the scale in the integrand of the cross-section is set by the charm quark mass for low virtualities $Q^2<m_c^2$.  The CGC band in \fig{f-vw1} also includes the uncertainties associated with choosing the charm mass within the range $m_c=1.2 \div 1.4$ GeV extracted from a global analysis of existing data at small-x $x<0.01$ \cite{ipsat-new,bcgc-new}.  In \fig{f-vw1}, we compare with the LHCb updated data released in 2014 \cite{lhcbn} which  are significantly more precise compared to earlier measurements \cite{lhcb}. It is seen that the ALICE and LHCb \cite{lhcbn} data are in good agreement with the CGC predictions while there seems to be some tensions between the experimental data and the pQCD results (labeled MNRT LO and NLO) at high $W_{\gamma p}$.  In \fig{f-vw1} (right panel), we show the charm-mass dependence of the total $J/\psi$ cross-section as a function of $W_{\gamma p}$.
It is seen that the combined ALICE and the LHCb updated 2014 data \cite{lhcbn} are more in favour of the saturation than of the 1-Pomeron model results at high $W_{\gamma p}$.

\begin{figure}[t]
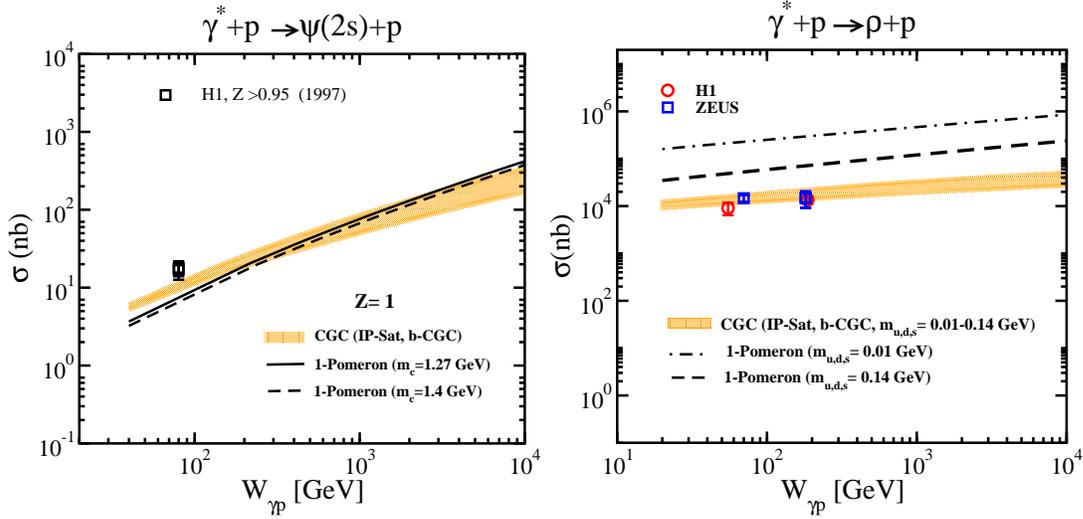
       
\includegraphics[width=0.47\textwidth,clip]{plot-psi-w-pho3.eps}
\includegraphics[width=0.47\textwidth,clip]{plot-rho-w-mass.eps}
\caption{Total $\psi(2s)$ and $\rho$ diffractive photo-production cross-section as a function of $W_{\gamma p}$, compared to results from the IP-Sat and the b-CGC  models with different quark masses. The results of the CGC/saturation (orange band) and 1-Pomeron models are also compared. The plots are taken from \cite{na-VM}. }
  \label{f-psi1}
\end{figure}  

\begin{figure}[t]
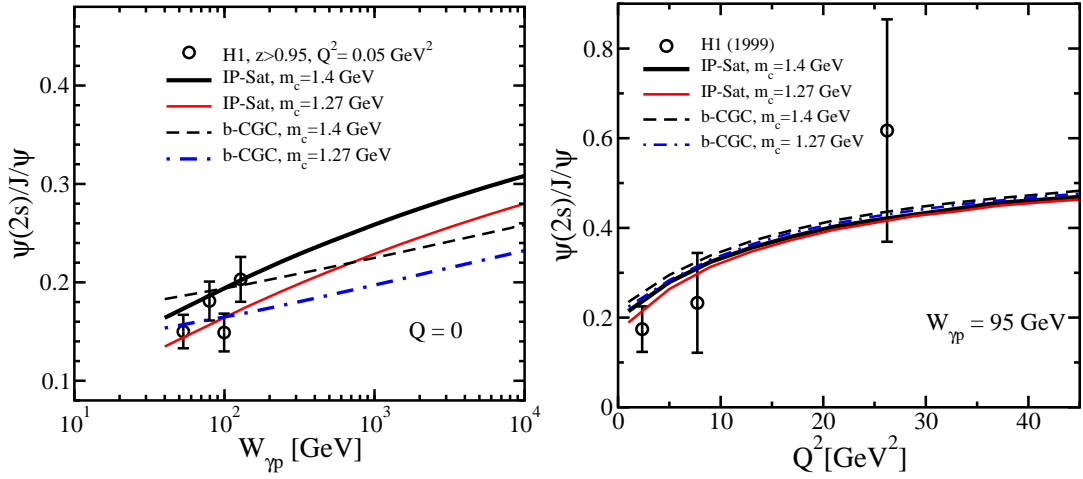
       
\includegraphics[width=0.47\textwidth,clip]{plot-psi-w-photo.eps}
\includegraphics[width=0.47\textwidth,clip]{plot-psi-q2-22.eps}
\caption{ The ratio of the cross-section for $\psi(2s)$ and $J/\psi$ diffractive production as a function of $W_{\gamma p}$ at $Q=0$ (left panel) and as a function of $Q^2$ at a fixed $W_{\gamma p}=95$ GeV (right panel). The plots are taken from Ref.\,\cite{na-VM}.}
  \label{f-ratio}
\end{figure}

In \fig{f-psi1} (left panel), we show total cross-section of elastic diffractive photoproduction of $\psi(2s)$ as a function of $W_{\gamma p}$ obtained from the IP-Sat and the b-CGC saturation models with different charm masses corresponding to different parameter sets of the dipole amplitude.  Note that the experimental data \cite{h1-psi} are for quasi-elastic ($Z>0.95$) photoproduction of $\psi(2s)$ while all theory curves are for elastic diffractive production with elasticity $Z=1$. It is seen that  within theoretical uncertainties associated with charm mass, the 1-Pomeron and the saturation models give rather similar results in the range of energy shown in \fig{f-psi1}. This is mainly due to the fact that the  $\psi(2s)$ is heavier than $J/\psi$, therefore effective dipole sizes which contribute to the total cross-section are smaller for $\psi(2s)$ than for $J/\psi$. Note that although the scalar part of the $\psi(2s)$ wave function extends to large dipole sizes, due to the existence of the node, there is large cancellation between dipole sizes above and below the node position. As a result, the total cross-section of $\psi(2s)$ is suppressed compared to $J/\psi$ production, see Figs.\,\ref{f-psi1},\ref{f-ratio}. 

In \fig{f-psi1} (right panel), we show total diffractive photoproduction of $\rho$ meson cross-section as a function of $W_{\gamma p}$, compared to results obtained from the b-CGC and the IP-Sat models. The orange band labeled CGC includes results from both the IP-Sat and  the b-CGC models with uncertainties associated to our freedom to choose different light-quark masses within a range $m_{u,d,s}= 0.01 \div 0.14$ GeV.  We also compare the CGC/saturation results with those obtained from the 1-Pomeron model with two different light quark masses  $m_{u,d,s}= 0.01$ and $0.14$ GeV.  It is seen that 1-Pomeron results are significantly different from the saturation models, and already HERA data can rule out the 1-Pomeron model with light quark masses. This may indicate the existence of large non-linear effects for the diffractive photoproduction of the $\rho$ meson.  Note that, as we already pointed out, the effective dipole size which contributes to the cross-section is proportional to the inverse of the meson mass at $Q=0$. Therefore, the total diffractive cross-section of lighter vector meson such as $\rho$ meson should be a better probe of saturation physics.

In \fig{f-ratio}, we show the ratio of the cross-section for $\psi(2s)$ and $J/\psi$ for diffractive production $R=\psi(2s)/J/\psi$ as functions of $W_{\gamma p}$ at $Q=0$ (left panel) and $Q^2$ at a fixed $W_{\gamma p}=95$ GeV (right panel).  It is seen that the ratio $R$ increases with virtualities at a fixed  $W_{\gamma p}$. 
It is also seen in  \fig{f-ratio} (left panel) that the photoproduction ratio $R (Q=0)$ increase with $W_{\gamma p}$ and becomes sensitive to different saturation models.  Therefore, precise measurements of the ratio of  diffractive photoproduction of $\psi(2s)$ and $J/\psi$ at HERA and the LHC can provide valuable extra constrain on the saturation models. We found that at fixed high virtualities, the ratio $R$  has little dependence to $|t|$ and $W_{\gamma p}$ (not shown here).

\begin{figure}[t]
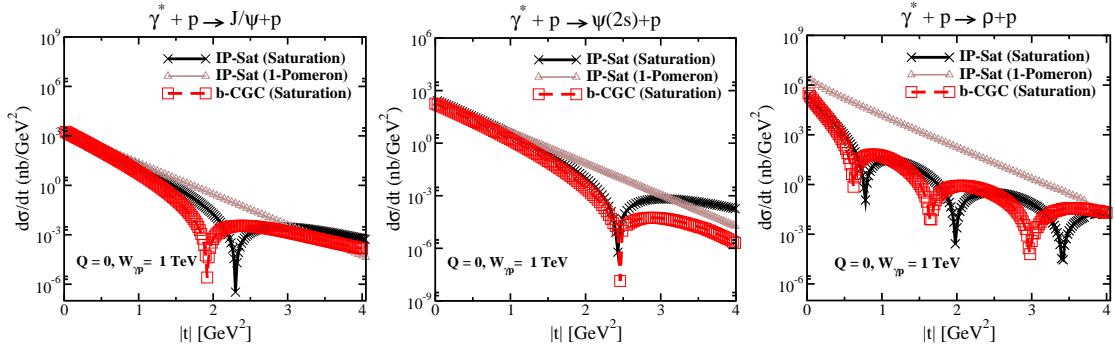
       
\includegraphics[width=0.32\textwidth,clip]{plot-jpsi-t-lhcb1.eps}
\includegraphics[width=0.32\textwidth,clip]{plot-psi-t-lhc.eps}
\includegraphics[width=0.32\textwidth,clip]{plot-rho-t-lhcb.eps}
\caption{ Differential diffractive vector meson photoproduction cross-sections for $J/\psi, \psi(2s)$ and $\rho$, as a function of $|t|$ within the IP-Sat (saturation), the b-CGC and the 1-Pomeron models at a fixed $W_{\gamma p}=1$ TeV and $Q=0$. The plots are taken from Ref.\,\cite{na-VM}. }
  \label{f-vt2}
\end{figure}

In \fig{f-vt2}, we compare the results obtained from the IP-Sat and the b-CGC models with those from the 1-Pomeron model, for the $t$-distribution of the elastic photoproduction of vector mesons $J/\psi$, $\psi(2s)$ and $\rho$ off the proton at an energy accessible at the LHC/LHeC, $W_{\gamma p}=1$ TeV, for $Q=0$.  Drastic different patterns for the diffractive $t$-distribution emerge between saturation and non-saturation models for lighter vector mesons production such as $\rho$ and $\phi$, with the appearance of multiple dips. 

The emergence of a single or multiple dips in the $t$-distribution of the vector mesons in the saturation models is directly related to the saturation (unitarity) features of the dipole scattering amplitude $\mathcal{N}$ at large dipole sizes.  In the 1-Pomeron model, since the impact-parameter profile of the dipole amplitude is a Gaussian for all values of $r$, its Fourier transform becomes exponential for all values of $t$. 
However,  in a case that the typical dipole size which contributes to the integral of cross-section is within the unitarity or black-disc limit, with $\mathcal{N}\to 1$,  the Fourier transform of the dipole amplitude in impact-parameter space leads to a dip or multi-dips. The saturation effect becomes more important at smaller Bjorken-$x$ or larger $W_{\gamma p}$, and lower virtualities $Q$ where the the contribution of large dipole sizes becomes more important.  For lighter vector mesons, the overlap extends to larger dipole sizes resulting in a  dip structure. In saturation models, the dips in the $t$-distribution recede towards lower $|t|$  with decreasing mass of the vector meson, increasing energy or decreasing  Bjorken-$x$, and decreasing virtuality $Q$. It is important to note that the main difference between a dipole model with linear and non-linear evolution (incorporating saturation effects through some specific model as those employed in this work) is that the former does not lead to the black-disc limit and, therefore, the dips do not systematically shift toward lower $|t|$ by increasing $W_{\gamma p}$, $1/x$, and $r$ or $1/Q$, while the latter does. Non-linear evolution evolves any realistic profile in $b$, like a Gaussian or Woods-Saxon distribution, and makes it closer to a step-like function in the $b$-space by allowing an increase in the periphery of the hadron (the dilute region) while limiting the growth in the denser centre.  This leads to the appearance of dips with non-linear evolution even if the dips were not present at the initial condition at low energies or for large $x$ (e.g. a Gaussian profile), or to the receding of dips towards lower values of $|t|$ even if they were already present in the initial condition (e.g. with a Woods-Saxon type profile).

To conclude: we showed that the recent LHC data on diffractive $J/\psi$ photoproduction are in good agreement with the saturation/CGC predictions while there are some tensions between recent LHCb and ALICE data with the 1-Pomeron model and pQCD results, see Fig.\,\ref{f-vw1}. This can be considered as the first hint of saturation effects at work in diffractive photoproduction of vector mesons off proton at the LHC. We showed that the photoproduction of $\rho$ meson is an excellent probe of small-x physics and already at HERA, 1-Pomeron model fails to describe the data with a reasonable light quark masses, see Fig.\,\ref{f-psi1}. We also provided predictions for the ratio of diffractive production of $\psi(2s)$ to $J/\psi$, namely $R=\psi(2s)/J/\psi$ at HERA and the LHC. We showed that while at high virtualities $R$ has little $|t|$ and $W_{\gamma p}$  dependence, it moderately increases with virtuality $Q$ at a fixed $W_{\gamma p}$.  We also found that the photoproduction ratio $R (Q=0)$ increases with $W_{\gamma p}$ and becomes  sensitive to different saturation models, see Fig.\,\ref{f-ratio}. Finally, We showed that the $t$-differential cross-section of exclusive production of vector mesons in high-energy collisions offers a unique opportunity to probe the saturation regime and discriminate between saturation and non-saturation models, see  Fig.\,\ref{f-vt2}.

\end{document}